\documentclass[11pt]{article}
\usepackage{amsfonts}
\usepackage{graphicx}
\usepackage{epstopdf}
\usepackage{epsfig}
\usepackage{amssymb}
\usepackage{setspace}
\usepackage{caption}
\usepackage{amsfonts}
\usepackage{color}
\usepackage{amsmath}
\usepackage{float}
\usepackage{comment}
\newcommand {\be}{\begin{equation}}
\newcommand {\ee}{\end{equation}}
\newcommand {\bea}{\begin{array}}
	
	\newcommand {\eea}{\end{array}}
\newcommand{\RN}{Reissner-Nordstrom~}

\textwidth=6.5in \hoffset=-.75in \voffset=-1in \numberwithin{equation}{section}
\numberwithin{figure}{section}

\begin{document}

	\begin{titlepage}
	\vspace{1cm} 
	\begin{center}
		{\Large \bf {Charged Taub-NUT-de Sitter spacetime in DGP braneworld and its thermodynamics}}\\
	\end{center}
	\vspace{2cm}
	\begin{center}
		\renewcommand{\thefootnote}{\fnsymbol{footnote}}
		Haryanto M. Siahaan{\footnote{haryanto.siahaan@unpar.ac.id}}\\
		Center for Theoretical Physics, Department of Physics,\\
		Parahyangan Catholic University,\\
		Jalan Ciumbuleuit 94, Bandung 40141, Indonesia
		\renewcommand{\thefootnote}{\arabic{footnote}}
	\end{center}
	\vspace{2cm}
\begin{abstract}
	We study a charged Taub-NUT spacetime solution in the DGP brane. We show that the Reissner-Nordstrom-Taub-NUT-de Sitter solution of Einstein-Maxwell gravity solves the corresponding equations of motion, where the cosmological constant is related to the cross-over scale in the DGP model. Following the approach by Teitelboim in discussing the thermodynamics of de Sitter spacetime and the proposal by Wu et al. for a conserved charge associated to the NUT parameter, we obtained the generalized Smarr mass formula and the first law of thermodynamics of the spacetime.
	
\end{abstract}

\end{titlepage}\onecolumn

\section{Introduction}
\label{sec:intro}

The view that the universe is a brane embedded in higher dimensional space has attracted great attentions in the last two decades. The absence of any experimental evidence for the existence of extra dimensions does not hinder the theoretical investigations in braneworld theories. Indeed, superstring theory which demands extra dimensions seems to be compatible with the braneworld proposal \cite{Nojiri:2003rz}. Moreover, the findings of gravitational waves and the black hole shadow have encouraged people to investigate these observations from the braneworld point of view \cite{Hou:2021okc,Eiroa:2017uuq,Dey:2020lhq,Toshmatov:2016bsb}. 

There are several braneworld gravity models in the literature, for example the Horava-Witten \cite{hor},
the Arkani Hamed-Dimopoulos-Dvali (ADD) \cite{ark}, the Randall-Sundrum (RS) \cite{ran}, and the Dvali-Gabadadze-Porrati (DGP) \cite{dgp} models. The latter, which is the topic investigated in this work, has several interesting properties in regards to the astronomical observations. It is well known that currently our universe is expanding, and it can be explained by the positive value of the cosmological constant in Einstein's field equations. The positive cosmological constant is natural in the DGP brane model, where it can be connected to a cross-scale in the theory that governs the transition from five- to four-dimension. In other words, the observed self accelerating universe is well understood in the DGP brane model. This explains the steady interest on this subject which can be seen from some recent works \cite{Gholami:2021olb,Iqbal:2019ooy,Chetry:2019fhz,Warkentin:2019caf,Biswas:2018mxe,Sbisa:2018ydq,Jawad:2018zeo}. In particular, some charged black hole solutions in the DGP model have been reported \cite{Chang-Young:2007iiv,Lee:2007nk,Lee:2008av}. 

In a model of gravity, in addition to the mass, charged, rotation, and even cosmological constant parameters, there can also exist the so-called NUT parameter \cite{Griffiths:2009dfa}. Despite some obstacles related to the spacetime with NUT parameter, such as the closed timelike curve and conic singularity, investigations about this spacetime can still be found recently in literature. For example, the collisional Penrose process in Kerr-Taub-NUT spacetime is studied in \cite{Zhou:2022eiv}, the thermodynamics related to NUT spacetime is explored in \cite{Liu:2022wku,Awad:2022jgn}, and even discussions for M5 branes on Taub-NUT related space in \cite{Gustavsson:2022jpo}. For the RS-II brane scenario, the related Taub-NUT spacetime has been discussed in \cite{Siahaan:2020bga}. It would be interesting to find a corresponding Taub-NUT related spacetime that fits in the DGP brane theory and the corresponding thermodynamics. The DGP brane spacetime solutions presented in \cite{Chang-Young:2007iiv,Lee:2007nk,Lee:2008av} have not incorporated the NUT parameter yet. 

The aim of this work is really to pursue this aim, namely finding an exact charged spacetime solution equipped with the NUT parameter which solve the corresponding Hamiltonian constraint in DGP model. The approach in solving the Hamiltonian constraint is similar to the works in refs. \cite{Dadhich,Aliev:2007fy,Siahaan:2020bga} in finding the black hole solution in RS-II brane scenario. First we employ the proper Kerr-Schild ansatz to solve the corresponding constraint, and then apply an appropriate coordinate transformation to get the Boyer-Lindquist form of the metric. It turns out that there are two cases of solutions for a charged Taub-NUT spacetime in DGP brane. The first case corresponds to the vanishing Ricci scalar, and the second one associates to a constant Ricci scalar. The latter case is just the \RN-Taub-NUT-de Sitter solution of Einstein-Maxwell theory. It is well known that studying the thermodynamic aspects of de Sitter spacetime is somehow delicate due to the existence of black hole and cosmological horizons in the spacetime. An interesting way to study the thermodynamics of a de Sitter black hole spacetime is proposed by Teitelboim \cite{Teitelboim:2002cv}, where we can have two different first laws of thermodynamics for each horizon. We can compute the Hawking temperature, entropy, angular velocity, and some other conjugate quantities for each horizon \cite{Sekiwa:2006qj}. As we discuss the thermodynamics regarding to the black hole horizon, we consider the cosmological horizon as a boundary where the incorporated physical parameters are set to be fixed. This is similar to the situation when one defines mass in ADM formalism where the boundary is at infinity. On the other hand, if we are investigating the thermodynamics that is related to the cosmological horizon, then the black hole horizon is considered as a boundary with fixed parameters.

Moreover, the spacetime we are studying in this work is equipped with NUT parameter as well. Thermodynamics of Taub-NUT spacetime is also another problem to be discussed. Some works in the past have been dedicated to it, for example the authors of \cite{BallonBordo:2019vrn} define the gravitational Misner charges to have the generalized first law in Taub-NUT spacetime.  The authors of \cite{Hennigar:2019ive} studied the thermodynamics of Taub-NUT-AdS spacetime and showed the entropy can be obtained by the Noether charge method. An interesting proposal is given in ref. \cite{Wu:2019pzr} where the authors define a new conserved quantity that associates to NUT parameter. The proposal of new conserved quantity $J_l = M l$ in ref. \cite{Wu:2019pzr} resembles the angular momentum $J=Ma$ for Kerr spacetime. By using the new conserved quantity, the authors managed to get the generalized Smarr mass formula for the spacetime, and the corresponding first law of thermodynamics as well. 

In this paper, after showing the \RN-Taub-NUT-de Sitter solution solve the Hamiltonian constraints in DGP brane, we study its thermodynamics. We adopt the approaches by Teitelboim \cite{Teitelboim:2002cv} and Sekiwa \cite{Sekiwa:2006qj} to deal with thermodynamics aspects of de Sitter spacetime, and the proposal by Wu et al. \cite{Wu:2019pzr} in discussing NUT parameter contribution. The organization of this paper is as the followings. In the section \ref{sec.2.effectiveeqtn3brane}, we review the DGP brane action and corresponding constraint equations. In section \ref{sec.3.TNDGPbrane}, we show that the Taub-NUT-de Sitter solution solves the equations of motion for the case with no electromagnetic fields on the brane. Section \ref{section.4.charged}, we extend the solution to a charged case where we show the \RN-Taub-NUT-de Sitter solution can solve the constraints exactly. Thermodynamics aspects of the brane is discussed in section \ref{section.5.thermo}, and we give conclusions in section \ref{sec.conclusion}. In this paper we used natural units $G = c = \hbar = 1$.

\section{Action and equation of motion}\label{sec.2.effectiveeqtn3brane}

In this chapter, we review the action and equations of motion associated with the DGP brane model. In the presence of sources, the DGP gravitational action takes the form \cite{dgp}
\be \label{action}
S = {\tilde M^3}\int {{d^5}x\sqrt { - \tilde g} } \tilde R + \int {{d^4}x\sqrt { - g} } \left( {M_P^2R + {L_{matter}}} \right)\,,
\ee 
where $\tilde R$ and $R$ are the five- and four-dimensional Ricci scalars, respectively. In the action above, $L_{matter}$ is a Lagrangian for the matters localized on the brane. The five dimensional spacetime coordinates are denoted by ${\tilde x}^M = \left(x^\mu,x^4=z\right)$, where $M = 0,1,2,3,4$, and $\mu = 0,1,2,3$. The determinants of five- and four-dimensional spacetime metric are given by $\tilde g$ and $g$, respectively, where the two metrics are related by $g_{\mu\nu} = {\tilde g}_{\mu \nu }\left(x^\mu,z=0\right) $. We define here the cross-over scale $\gamma = 1/{\lambda} = M_P^2/2{\tilde M}^3$. Moreover, the boundary is at $z=0$ and we assume the ${\mathbb Z}_2$ symmetry across boundary. Varying the action (\ref{action}) with respect to the tensor metric $\tilde{g}_{MN}$ yields the equation of motion
\be \label{Einstein.bulk}
{{\tilde G}_{MN}} = {{\tilde \kappa }^2}\sqrt {\frac{g}{{\tilde g}}} \delta _M^\mu \delta _N^\nu \left( {{T_{\mu \nu }} - {\kappa ^{-2}}{G_{\mu \nu }}} \right)  \delta\left(z\right)\,,
\ee 
where ${\tilde G}_{MN} = {\tilde R}_{MN} - \tfrac{1}{2}{\tilde g}_{MN} {\tilde R}$ is the five-dimensional Einstein tensor, $G_{\mu \nu }$ is the four-dimensional Einstein tensor, $R_{MN}$ is the five-dimensional Ricci tensor, $\kappa^2 = 1/M_P^2$, and ${\tilde \kappa}^2 = 1/{\tilde M}^3$. In the equation above, $T_{MN}$ is the energy-momentum tensor in the bulk, whereas $T_{\mu\nu}$ is the energy-momentum tensor for matters localized on the brane. 

We decompose the bulk metric into the following form
\be
ds^2  = \tilde g_{MN} d\tilde x^M d\tilde x^N  = {\tilde g}_{\mu \nu } \left( {x,z} \right)dx^\mu  dx^\nu   + 2{\cal N}_\mu  dx^\mu  dz
+ \left( {{\cal N}^2  + {\tilde g}_{\mu \nu } {\cal N}^\mu  {\cal N}^\nu  } \right)dz^2 \,.
\ee 
Accordingly, ${\tilde G}_{\mu z}$ and ${\tilde G}_{z z}$ components of (\ref{Einstein.bulk}) give the following equations \cite{Aliev:2004ds}
\be \label{eq.mom}
{\tilde{\nabla}} _\alpha  K^\alpha  _\mu   - {\tilde{\nabla}} _\mu  K = 0\,,
\ee 
and
\be \label{eq.Ham}
R - K^2  + K_{\mu \nu } K^{\mu \nu }  = 0\,.
\ee 
The extrinsic curvature tensor $K_{\mu\nu}$ above is given by
\be 
K_{\mu \nu }  = \frac{1}{{2{\cal N}}}\left( {\partial _z {\tilde g}_{\mu \nu }  - {\tilde{\nabla}} _\mu  {\cal N}_\nu   - {\tilde{\nabla}} _\nu  {\cal N}_\mu  } \right)\,,
\ee 
and ${\tilde{\nabla}} _\mu$ is the covariant derivative associated to the metric tensor ${\tilde g}_{\mu\nu}$.
The two equations (\ref{eq.mom}) and (\ref{eq.Ham}) above are known as the momentum and Hamiltonian constraint equations, respectively. Moreover, the corresponding Israel's junction condition with ${\mathbb Z}_2$ symmetry can be obtained by integrating both sides of eq. (\ref{Einstein.bulk}) along the $z$ direction and followed by taking the limit $z\to 0$. It reads
\be \label{Israel}
G_{\mu \nu }  = \kappa ^2 T_{\mu \nu }  + {\lambda} \left( {K_{\mu \nu }  - g_{\mu \nu } K} \right)\,.
\ee 
If one considers the traceless $T_{\mu\nu}$ condition associated to the electromagnetic field trapped on the brane, eq. (\ref{Israel}) yields the momentum constraint (\ref{eq.mom}) to be satisfied \cite{Lee:2007nk}, while the Hamiltonian condition gives
\be \label{Ham}
R_{\mu \nu } R^{\mu \nu }  + {\lambda}^2 R -\frac{R^2}{3}  + \kappa ^4 T_{\mu \nu } T^{\mu \nu }  - 2\kappa ^2 R_{\mu \nu } T^{\mu \nu }  = 0\,.
\ee 
The equation of motion on the brane can be obtained by using the Israel's junction condition (\ref{Israel}) into the Einstein equations in the bulk, i.e. $z \ne 0$, which reads
\[
R_{\mu \nu }  - \frac{1}{2}g_{\mu \nu } R + E_{\mu \nu }  =  - \frac{{\kappa ^4 }}{{{\lambda}^2 }}\left( {T_\mu ^\alpha  T_{\alpha \nu }  - \frac{1}{2}g_{\mu \nu } T_{\alpha \beta } T^{\alpha \beta } } \right) 
\]
\[
- \frac{1}{{{\lambda}^2 }}\left( R_\mu ^\alpha  R_{\alpha \nu }  - \frac{2}{3}RR_{\mu \nu }+ \frac{1}{4}g_{\mu \nu } R^2  - \frac{1}{2}g_{\mu \nu } R_{\alpha \beta } R^{\alpha \beta }  \right)
\]
\be \label{EinsteinEqtn}
+ \frac{{\kappa ^2 }}{{{\lambda}^2 }}\left( {R_\mu ^\alpha  T_{\alpha \nu }  + T_\mu ^\alpha  R_{\alpha \nu }  - \frac{2}{3}RT_{\mu \nu }  - g_{\mu \nu } R_{\alpha \beta } T^{\alpha \beta } } \right)\,.
\ee 
In the equation above, $E_{\mu\nu}$ is the traceless ``electric part'' of the five-dimensional Weyl tensor $C_{KLMN}$, and we have set ${\lambda} = 2\kappa^2/{\tilde \kappa}^2$ and considered $\kappa^2 = 8\pi$. 

\section{Taub-NUT DGP brane}\label{sec.3.TNDGPbrane}

Let us first consider the vacuum case on the brane, namely $T_{\mu\nu} = 0$. It yields the Hamiltonian constraint (\ref{Ham}) reduces to
\be \label{Ham.vacuum}
R_{\mu \nu } R^{\mu \nu }  - \frac{1}{3}R^2  + {\lambda}^2 R = 0\,.
\ee 
It turns out that this equation can be satisfied in two cases. The first one, which will be referred as the flat case, corresponds to the vanishing of four-dimensional Ricci scalar and squared Ricci tensor,
\be \label{Ham.vacuumFlat}
R = 0~~~,~~~R_{\mu\nu}R^{\mu\nu}=0\,.
\ee 
The second case, i.e. the non-flat one, the related Ricci scalar and squared Ricci tensor satisfy
\be \label{Ham.vacuumNonFlat}
R = 12 {\lambda}^2~~,~~R_{\mu\nu}R^{\mu\nu} = 36 {\lambda}^4\,,
\ee 
respectively.

In solving the Hamilton constraint condition (\ref{Ham.vacuumFlat}), let us use the Kerr-Schild form for the metric
\be \label{KSmetric}
{\rm{d}}s^2  = {\rm{d}}s_{\rm flat}^2  + H\left( {r,x} \right)\left( {k_\mu  {\rm{d}}x^\mu  } \right)^2 
\ee 
where
\[
{\rm{d}}s_{\rm flat}^2  =  - \left( {\frac{{r^2  - l^2 }}{{r^2  + l^2 }}} \right)\left( {{\rm{d}}u^2  + 4lx{\rm{d}}\psi {\rm{d}}u} \right)
+ \frac{{\Delta _x {\left( {r^2  + l^2 } \right)^2  - 4l^2x^2 \left( {r^2  - l^2 } \right)}}}{{r^2  + l^2 }}{\rm{d}}\psi ^2 
\]
\be 
+ \frac{{r^2  + l^2 }}{{\Delta _x }}{\rm{d}}x^2  + 2{\rm{d}}u{\rm{d}}r + 4lx{\rm{d}}\psi {\rm{d}}r\,,
\ee
$\Delta_x = 1-x^2$, and $k_\mu {\rm d}x^\mu = {\rm d}u + 2 lx {\rm d}\psi$. Here, $l$ denotes the NUT parameter. Accordingly, the corresponding Ricci scalar and squared of Ricci tensor related to the metric (\ref{KSmetric}) can be expressed as
\be \label{Ricci.KS}
R = \frac{{\partial ^2 H\left( {r,x} \right)}}{{\partial r^2 }} + \frac{{4r}}{{\left( {r^2  + l^2 } \right)}}\frac{{\partial H\left( {r,x} \right)}}{{\partial r}} + \frac{{2H\left( {r,x} \right)}}{{\left( {r^2  + l^2 } \right)}}\,,
\ee 
and
\[
R_{\mu \nu } R^{\mu \nu }  = \frac{1}{2}\left( {\frac{{\partial ^2 H\left( {r,x} \right)}}{{\partial r^2 }}} \right)^2 
+ \frac{2}{{\left( {r^2  + l^2 } \right)^2 }}\left[ r\left( {r^2  + l^2 } \right)\frac{{\partial H\left( {r,x} \right)}}{{\partial r}}   + 2l^2 H\left( {r,x} \right) \right]\frac{{\partial ^2 H\left( {r,x} \right)}}{{\partial r^2 }}
\]
\be \label{RicciTensor2KS}
+ \frac{{4r^2 }}{{\left( {r^2  + l^2 } \right)^2 }}\left( {\frac{{\partial H\left( {r,x} \right)}}{{\partial r}}} \right)^2 + \frac{{4rH\left( {r,x} \right)}}{{\left( {r^2  + l^2 } \right)^2 }}\frac{{\partial H\left( {r,x} \right)}}{{\partial r}} + \frac{{2\left( {r^4  + 5l^4  - 2l^2 r^2 } \right)}}{{\left( {r^2  + l^2 } \right)^4 }}H\left( {r,x} \right)^2 \,,
\ee
respectively. The solution for $H\left(r,x\right)$ that satisfies both equations in (\ref{Ham.vacuumFlat}) is given by\footnote{Note that the vanishing of squared Ricci tensor $R_{\mu\nu}R^{\mu\nu} = 0$ demands the absence of ``tidal charge'' $\beta$, which exists for the black hole in RS braneworld \cite{Chamblin}.}
\be \label{Hsol}
H\left( {r,x} \right) = \frac{{2Mr}}{{r^2  + l^2 }}\,.
\ee 

Now let us obtain the Boyer-Lindquist form of the metric (\ref{KSmetric}) with eq. (\ref{Hsol}) as the $H\left(r,x\right)$ function. We can employ the following transformation
\be 
{\rm d}u = {\rm d}t + \frac{{r^2  + l^2 }}{{r^2  - 2Mr - l^2 }}{\rm d}r~~,~~{\rm d}\psi  = {\rm d}\phi\,, 
\ee 
to the line element (\ref{KSmetric}) and the result can be written as
\[
{\rm d}s^2  =  - \frac{{r^2  - 2Mr - l^2 }}{{r^2  + l^2 }}\left( {{\rm d}t + 2lx{\rm d}\phi } \right)^2 
\]
\be \label{Taub-NUT}
+ \left( {r^2  + l^2 } \right)\left( {\frac{{{\rm d}r^2 }}{{r^2  - 2Mr - l^2 }} + \frac{{{\rm d}x^2 }}{{\Delta _x }}} \right) + \left( {r^2  + l^2 } \right)\Delta _x {\rm d}\phi ^2 \,.
\ee 
The last equation is the well known Taub-NUT metric \cite{Griffiths:2009dfa} which solves the vacuum Einstein equation. However, it is obvious that Taub-NUT solution solves the constraint (\ref{Ham.vacuumFlat}), since vacuum Einstein requires all the components of Ricci tensor to be zero and obviously lead to the vanishing Ricci scalar. 

For the non-flat case with a set of constraint equations given in (\ref{Ham.vacuumNonFlat}), we can employ the Kerr-Schild metric as appear in (\ref{KSmetric}) with ${\rm d}s_{\rm flat}^2 \to {\rm d}s_{\rm non-flat}^2$ replacement, i.e.
\be \label{KSmetric.nonflat}
{\rm{d}}s^2  = {\rm{d}}s_{\rm non-flat}^2  + H\left( {r,x} \right)\left( {k_\mu  {\rm{d}}x^\mu  } \right)^2 \,,
\ee 
where
\[
{\rm{d}}s_{{\rm{non - flat}}}^{\rm{2}}  ={2{\rm{d}}u{\rm{d}}r + 4lx{\rm{d}}r{\rm{d}}\psi } + \frac{{\left( {r^2  + l^2 } \right)}}{{\Delta _x }}{\rm{d}}x^2  - \frac{Z}{{\left( {r^2  + l^2 } \right)}}{\rm{d}}u^2  
\]
\be \label{sol.nonflat}
+ \frac{{\Delta _x \left( {r^2  + l^2 } \right)^2  - 4l^2 x^2 Z}}{{\left( {r^2  + l^2 } \right)}}{\rm{d}}\psi ^2 - \frac{{4lxZ}}{{\left( {r^2  + l^2 } \right)}}{\rm{d}}u{\rm{d}}\psi \,,
\ee
and 
\be \label{Zsol}
Z = r^2 - l^2 - {\lambda}^2\left(r^4 + 6 r^2 l^2 -3l^4\right).
\ee 
It turns out that this metric gives the Ricci scalar and squared of Ricci tensor exactly as indicated in (\ref{Ham.vacuumNonFlat}) for $H\left(r,x\right)$ as appeared in (\ref{Hsol}). To bring the line element (\ref{KSmetric.nonflat}) into the Boyer-Lindquist expression, one can employ the coordinate transformation
\be 
{\rm{d}}u = {\rm{d}}t + \frac{{r^2  + l^2 }}{{Z - 2Mr}}{\rm{d}}r~~,~~{\rm{  d}}\psi  = {\rm{d}}\phi \,,
\ee 
and the result is
\be 
{\rm{d}}s^2  =  - \frac{{Z - 2Mr}}{{\left( {r^2  + l^2 } \right)}}\left( {dt + 2lxd\phi } \right)^2
+ \left( {r^2  + l^2 } \right)\left( {\frac{{dr^2 }}{{Z - 2Mr}} + \frac{{dx^2 }}{{\Delta _x }}} \right) + \left( {r^2  + l^2 } \right)\Delta _x d\phi ^2 \,.
\ee 
The last equation can be understood as the Taub-NUT-de Sitter spacetime \cite{Griffiths:2009dfa} with cosmological constant $\Lambda = 3 {\lambda}^2$. Obviously, this is a straightforward generalization of the solution, from the flat case obeying the conditions in (\ref{Ham.vacuumFlat}) to the non-flat one that satisfies eq. (\ref{Ham.vacuumNonFlat}). Alternatively, one can use the ansatz for flat case (\ref{KSmetric}) directly to solve (\ref{Ham.vacuumNonFlat}), where the result for $H\left(r,x\right)$ function is
\be \label{Hsol.vacuum.nonflat}
H\left( {r,x} \right) = \frac{{2Mr + {\lambda}^2 \left( {r^4  + 6l^2 r^2  - 3l^4 } \right)}}{{\left( {r^2  + l^2 } \right)}}\,.
\ee
Plugging (\ref{Hsol.vacuum.nonflat}) into eq. (\ref{KSmetric}) yields exactly the same metric as appeared in eq. (\ref{KSmetric.nonflat}) with the $H\left(r,x\right)$ function as given in (\ref{Hsol}).

\section{Charged solution}\label{section.4.charged}

After discussing some neutral solutions in the previous section, let us now turn to the electrically charged cases. Here we consider the existence of source-free Maxwell fields outside the Taub-NUT black hole localized on the DGP brane obeying the equations of motion
\be 
\nabla _\mu  F^{\mu \nu }  = 0\,,
\ee 
and the Bianchi identity $\nabla _{[\alpha } F_{\mu \nu ]}  = 0$. Note that the covariant derivative $\nabla_\mu$ is defined with respect to the tensor metric $g_{\mu\nu}$ on the brane. In the presence of Maxwell fields, i.e. $T_{\mu\nu} \ne 0$, solving the Hamiltonian constraint (\ref{Ham}) can be done by considering
\[
R = 0~,~R_{\alpha \beta } R^{\alpha \beta }  = \frac{{4Q^4 }}{{\left( {r^2  + l^2 } \right)^4 }}~~,
\]
\be \label{Ham.Maxwell.flat}
R_{\mu \nu } T^{\mu \nu }  = \frac{{Q^4 }}{{2\pi \left( {r^2  + l^2 } \right)^4 }}~,~T_{\mu \nu } T^{\mu \nu }  = \frac{{Q^4 }}{{16\pi ^2 \left( {r^2  + l^2 } \right)^4 }}\,,
\ee 
for the flat case, and
\[
R = 12{\lambda}^2 ~~,~~R_{\mu \nu } T^{\mu \nu }  = \frac{{Q^4 }}{{2\pi \left( {r^2  + l^2 } \right)^4 }}~,
\]
\be \label{Ham.Maxwell.nonflat}
R_{\alpha \beta } R^{\alpha \beta }  = 36{\lambda}^4  + \frac{{4Q^4 }}{{\left( {r^2  + l^2 } \right)^4 }}~,~T_{\mu \nu } T^{\mu \nu }  = \frac{{Q^4 }}{{16\pi ^2 \left( {r^2  + l^2 } \right)^4 }}\,,
\ee 
for the non-flat one. Recall that the energy-momentum tensor related to the Maxwell fields is given by
\be 
T_{\mu \nu }  = \frac{1}{{4\pi }}\left( {F_{\mu \alpha } F_\nu ^\alpha   - \frac{1}{4}g_{\mu \nu } F_{\alpha \beta } F^{\alpha \beta } } \right)\,.
\ee 

For the flat case, we find the metric ansatz (\ref{KSmetric}) with
\be\label{H.Maxwell}
H\left(r,x\right) = \frac{2Mr-Q^2}{\left(r^2+l^2\right)}
\ee 
accompanied by the vector
\be \label{Amu}
A_\mu  {\rm{d}}x^\mu   =  - \frac{{Qr}}{{r^2  + l^2 }}\left( {{\rm{d}}u + 2lx{\rm{d}}\psi } \right)\,,
\ee 
solve the constraints appeared in (\ref{Ham.Maxwell.flat}). Again, the solution differs to that of RS-II brane \cite{Siahaan:2020bga} family of black hole solution, where it does not incorporate the ``tidal charge'' which is interpreted as the extra dimensional effect in RS-II brane \cite{Dadhich}. On the other hand, accompanied by the vector solution (\ref{Amu}), the non-flat constraints (\ref{Ham.Maxwell.nonflat}) can be satisfied by the metric ansatz (\ref{KSmetric.nonflat}) with the corresponding $H\left(r,x\right)$ function given in eq. (\ref{H.Maxwell}). The corresponding Boyer-Lindquist form can be achieved by using the coordinate transformation
\be 
{\rm{d}}u = {\rm{d}}t + \frac{{r^2  + l^2 }}{{\Delta_Q }}{\rm{d}}r~~,~~{\rm{d}}\psi  = {\rm{d}}\phi \,,
\ee 
where $\Delta_Q = Z - 2Mr + Q^2$  and the $Z$ function is given in (\ref{Zsol}). Accordingly, the resulting line element for a charged spacetime on the brane in non-flat case can be written as
\be \label{RNTNdSmetric}
{\rm{d}}s^2  =  - \frac{{\Delta _Q }}{{\left( {r^2  + l^2 } \right)}}{\rm{d}}t^2  + \frac{{\left( {r^2  + l^2 } \right)}}{{\Delta _Q }}{\rm{d}}r^2  + \frac{{\left( {r^2  + l^2 } \right)}}{{\Delta _x }}{\rm{d}}x^2 - \frac{{4lx\Delta _Q }}{{\left( {r^2  + l^2 } \right)}}{\rm{d}}t{\rm{d}}\phi  + \frac{{\left( {r^2  + l^2 } \right)^2 \Delta _x  - 4l^2 x^2 \Delta _Q }}{{\left( {r^2  + l^2 } \right)}}{\rm{d}}\phi ^2 \,,
\ee 
which is just the Reissner-Nordstrom-Taub-NUT-de Sitter (RNTNdS) solution of Einstein-Maxwell theory with cosmological constant $\Lambda = 3 {\lambda}^2$. The metric (\ref{RNTNdSmetric}) is accompanied by the vector solution (\ref{Amu}) in solving the corresponding non-flat charged Hamiltonian constraints (\ref{Ham.Maxwell.nonflat}). 

So far we have showed that the resulting spacetime metric on the brane is just the family of RNTNdS type solution in Einstein-Maxwell theory with cosmological constant. However, in the corresponding equation of motion (\ref{EinsteinEqtn}), there exist the second rank tensor $E_{\mu\nu}$ which comes from projection of five-dimensional Weyl tensor $C_{ABCD}$ on the brane \cite{Shiromizu,Aliev:2004ds}, i.e. $E_{\mu \nu }  = C_{ABCD} n^A n^C e_\mu ^B e_\nu ^D $. It turns out that this ``electric part'' of five-dimensional Weyl tensor related to the charged solution above appears quite simple, where the non-zero parts can be obtained as
\be 
E_{tt} = -\frac{Q^2 \Delta_Q }{\left(r^2+l^2\right)^3}\,,
\ee 
\be 
E_{t\phi} = -\frac{2lx Q^2 \Delta_Q }{\left(r^2+l^2\right)^3}\,,
\ee 
\be 
E_{rr} = \frac{Q^2 }{\Delta_Q \left(r^2+l^2\right)}\,,
\ee 
\be 
E_{xx} = -\frac{Q^2 }{\Delta_x \left(r^2+l^2\right)}\,,
\ee 
\be 
E_{\phi \phi }  =  - \frac{{Q^2 \left( {\left( {r^2  + l^2 } \right)^2 \Delta _x  + 4l^2 x^2 \Delta _Q } \right)}}{{\left( {r^2  + l^2 } \right)^3 }}\,.
\ee 
By assuming the existence of this second rank tensor $E_{\mu\nu}$, one can accept that the above metric and vector solution obey the Einstein equations (\ref{EinsteinEqtn}).

Obviously the solutions presented here and in the last section solve the Einstein equations
\be \label{Einstein.evac.eq}
R_{\mu \nu }  - \Lambda g_{\mu \nu }  = 2F_{\mu \alpha } F_\nu ^\alpha   - \frac{1}{2}g_{\mu \nu } F_{\alpha \beta } F^{\alpha \beta } \,.
\ee 
Provided that the corresponding $E_{\mu\nu}$ does exist, these solutions can also solve the Einstein equations (\ref{EinsteinEqtn}) and the Hamiltonian constraint (\ref{Ham}) as well. In fact, every solution of eq. (\ref{Einstein.evac.eq}) solve the Hamiltonian constraint (\ref{Ham}) and moreover the Einstein equations on the brane (\ref{EinsteinEqtn}) by assuming the existence of associated $E_{\mu\nu}$.

\section{Thermodynamics of the non-flat case}\label{section.5.thermo}

In previous section, we have shown that the de Sitter family spacetime with $\Lambda = 3 {\lambda}^2$ is a generic solution in the DGP brane for the non-flat case. A positive cosmological constant in the non-flat case can be explained from the cross-over scale between the bulk and the brane in DGP theory. In other words, the expansion of our universe can be considered as some extra-dimensional effects according to this point of view. Now in this section let us study the thermodynamics that associate to a charged object with NUT parameter in the non-flat case of the DGP brane, i.e. the thermodynamics of RNTNdS spacetime. The line element under consideration is the one appears in eq. (\ref{RNTNdSmetric}). As typical for a black hole in de Sitter geometry, there exist four zeroes of $\Delta_Q$ in eq. (\ref{RNTNdSmetric}) which are denoted by
\be 
r_{ -  - }  < r_ -   < r_ +   < r_{ +  + } \,.
\ee 
The largest root $r_{++}$ is acknowledged as the cosmological horizon $r_c$, whereas the smaller one $r_+$ is identified as the black hole horizon $r_h$. The existence of these two horizons is origin of the problem in formulating the thermodynamical aspects of the spacetime. Each horizon associates to a different temperature, and the two systems are not in thermal equilibrium. Furthermore, unlike in an asymptotically flat spacetime, there is no notion of observer at spatial infinity inside the cosmological horizon. Hence, defining the conserved quantities as the corresponding thermodynamical parameters requires some special methods. 

Here we adopt the approach presented in refs. \cite{Teitelboim:2002cv,Gomberoff:2003ea,Sekiwa:2006qj} to investigate the thermodynamics of black hole in the de Sitter background. In  this approach, as one discusses the thermodynamics that corresponds to the black hole horizon, the cosmological horizon is considered just as a boundary. It resembles the approach in an asymptotically flat black hole spacetime,where the ADM mass of black hole is measured by an observer at the boundary, i.e. at infinity. However, outside a black hole in de Sitter spacetime, there is no notion of spatial infinity since this point is located beyond the cosmological horizon \cite{Teitelboim:2002cv,Sekiwa:2006qj}. On the other hand, as one studies the thermodynamical aspects related to the cosmological horizon, the black hole horizon is considered as the boundary. Reversal of these horizons yields the sign changes of conserved quantities associated to the first law of thermodynamics with respect to each horizon, as we will see next.

Related to the black hole horizon thermodynamics, the conserved mass and electric charge in the spacetime with line element (\ref{RNTNdSmetric}) are given by
\be 
M_h = M~~,~~Q_h = Q\,.
\ee 
Above, the subscript ``$h$'' refers to the quantities related to the black hole horizon. On the other hand, the mass and electric charge that associates to the cosmological horizon are given by
\be 
M_c = - M~~,~~Q_c = -Q\,,
\ee 
where the subscript ``c'' stands for the parameters that correspond to the cosmological horizon. Indeed, in classical Einstein gravity, the cosmological constant is just a constant, and one cannot consider it to be varied in a thermodynamical relation. However, as argued in refs. \cite{Caldarelli:1999xj,Kubiznak:2015bya}, cosmological constant can be viewed as some thermodynamical variables in a semiclassical approach. Here we treat cosmological constant as a variable in thermodynamics as well and define the physical cosmological constant parameters as \cite{Sekiwa:2006qj}
\be 
\Lambda _h  = 3{\lambda}^2 ~~,~~\Lambda _c  =  - 3{\lambda}^2 \,.
\ee 

Intuitively, $\Lambda _h $ corresponds to the black hole horizon and $\Lambda _c $ associates to the cosmological one. In addition to the parameters above, we add another conserved quantities in the system which are related to the NUT parameter. Following prescriptions in ref. \cite{Wu:2019pzr}, we define
\be \label{NUTconserved}
N_h = M l ~~,~~N_c = -Ml\,, 
\ee 
as new quantities that associate to black hole and cosmological horizons, respectively. The definition of these new quantities is analogous to the conserved angular momentum in Kerr spacetime, i.e. $J= Ma$, where the NUT parameter $l$ plays a similar role as the rotational parameter $a$ in rotating spacetime. Clearly, here we do not provide any surface integrals to obtain the conserved quantities appeared in (\ref{NUTconserved}). Indeed there exist several works that can be used to construct such integral to obtain a charge that associates to NUT parameter, for example the covariant phase space approach \cite{Lee:1990nz,Wald:1993nt}. Particularly for the asymptotically locally de Sitter spacetimes, discussions on conservative charges can be found in refs. \cite{Anninos:2010zf,Compere:2020lrt,Kolanowski:2020wfg,Poole:2021avh}. Despite the absence of a surface integral formula for the conservative charge related to NUT parameter, the authors of \cite{Wu:2019pzr} have shown that defining $N=Ml$ leads to a nice generalized Smarr mass formula and furthermore can produce the expected entropy as a quarter of black hole area. Moreover, some previous works support this proposal, for example as pointed out by the authors of \cite{Mann:2005cx} where $N = Ml \equiv M_5$ can be viewed as a conserved mass of a five-dimensional magnetic monopole. Another supporting result is the work presented in ref. \cite{Aliev:2007fy} where the definition $N = Ml$ can explain the gyromagnetic ratio of Kerr-Taub-NUT spacetime. Therefore, we can expect the definition in (\ref{NUTconserved}) could give us some good results even though we now have both black hole and cosmological horizons instead of only the black hole one as in ref. \cite{Wu:2019pzr}.

In addition the quantities defined in eq. (\ref{NUTconserved}) which are treated as some thermodynamical variables, the NUT parameter $l$ itself should be regarded as another variable in thermodynamics as well to yield a consistent Bekenstein-Smarr mass formula. The Misner string attached at the south and north poles that can carry some rotation-like and electromagnetic-like energies \cite{Wu:2019pzr}. It is well known that the first law of thermodynamics for Kerr-Newman black hole can be written as
\be \label{KNthermo}
dM = TdS + \Omega_h dJ + \Phi_h dQ
\ee 
where $M$ is the black hole mass, $T$ is the Hawking temperature, $S$ is the black hole entropy, $\Omega_h$ is the angular velocity at horizon, $J$ is the angular momentum, $Q$ is the black hole charge, and $\Phi_h$ is electric potential at horizon. In the last equation, the change of rotational energy of Kerr-Newman black hole is represented by the term $\Omega_h dJ$, whereas the contribution of electromagnetic energy is given by the term $\Phi_h dQ$. 

In this work, we consider the first law of thermodyanamics for the Taub-NUT spacetime takes the similar form as in eq. (\ref{KNthermo}), namely
\be \label{TNthermo}
dM = TdS + \omega_h dN + \varpi_h dl\,,
\ee 
where the change of rotation-like energy is given by the term $\omega_h dN$ and the electromagnetic-like energy contribution can be found in the term $\varpi_h dl$. The definition of $N=Ml$ which resembles the angular momentum $J = Ma$ and the fact where NUT parameter $l$ is viewed as some conserved quantity just like $Q$ in Kerr-Newman case make the correspondence between eqs. (\ref{KNthermo}) and (\ref{TNthermo}) is clear. In Taub-NUT spacetime, we have the NUT potential at horizon $\varpi_h$ which acts like the electric potential at horizon $\Phi_h$ in Kerr-Newman spacetime. Later it can be shown that the first law of thermodynamics that takes the form (\ref{TNthermo}) can produce to the mass formula 
\be 
M = 2TS + 2\omega N + \varpi l\,.
\ee 

However, we have some subtleties here that are related to the NUT parameter as a conserved quantity. Indeed, so the NUT parameter $l$ can act exactly as the electric charge of a body in de Sitter spacetime, we should have $l_h = - l_c = l$. However, this setting leads to an inconsistency in the definition of conserved quantities $N = Ml$ with respect to black hole and cosmological horizons as given in (\ref{NUTconserved}). Recall that the quantity $N=Ml$ is inspired by the angular momentum $J = Ma$, and for a rotating object in de Sitter spacetime we have the relation $J_h = -J_c = Ma$ \cite{Sekiwa:2006qj}. Here we maintain this resemblance between $N$ and $J$, i.e. they both change sign as one reverses the horizons. Therefore, we need to consider the NUT parameter as a conserved quantity does not change sign as the horizons are reversed in their role, i.e.
\be \label{NUTrel}
l_h  = l = l_c\,.
\ee 
This is similar to the invariance of rotational parameter $a$ under the change of horizon's role in Kerr-de Sitter spacetime \cite{Sekiwa:2006qj}. 

\subsection{Black hole horizon thermodynamics}

In this section, we study the thermodynamics of black hole horizon where the incorporated physical parameters are
\be 
M_h = M~~,~~N_h = Ml~~,~~Q_h=Q~~,~~l_h=l~~,~~\Lambda_h = 3{\lambda}^2\,.
\ee 
The area of RNTNdS horizon is $A_h = 4\pi \Xi_h$ where $\Xi_h = r_h^2 +l^2$. From the condition $\Delta_Q\left(r_h\right)=0$, we can have the following relation
\be 
\Xi _h  = 2M_h r_h  + 2l^2  - Q_h^2  + {\lambda}^2 \left( {r_h^4  + 6l^2 r_h^2  - 3l^4 } \right)\,.
\ee 
Furthermore, the last equation can be rearranged as
\be \label{lambda14h}
\left( {1 - 4{\lambda}^2 l^2 } \right)\left( {\Xi _h  - 2l^2 } \right) + Q_h^2  - {\lambda}^2 \Xi _h^2  = 2M_h r_h\,.
\ee 
It turns out that adding $N_h^2$ to the squared of eq. (\ref{lambda14h}) can give the squared mass formula
\be \label{MassRelh}
4M_h^2 \Xi _h  = 4N_h^2  + \left\{ {\left( {1 - 4{\lambda}^2 l^2 } \right)\left( {\Xi _h  - 2l^2 } \right) + Q^2  - {\lambda}^2 \Xi _h^2 } \right\}^2 \,,
\ee 
which can be rewritten into the following form
\be \label{eq.Smarr2}
M_h^2  = \frac{1}{{4\Xi _h }}\left\{ \left( {1 - \frac{{32\pi }}{3}{\cal P}_h l^2 } \right)\left( {\Xi _h  - 2l^2 } \right) + Q_h^2  - \frac{{8\pi }}{3}{\cal P}_h \Xi _h^2  \right\}^2  + \frac{{N_h^2 }}{{\Xi _h }}\,.
\ee 
The last expression is the generalized Smarr formula that associates to the black hole horizon. In the equation above, we consider the cosmological constant $\Lambda_h = 3 \lambda^2$ as a thermodynamical variable which consequently yields the generalized pressure ${\cal P}_h= \Lambda_h \left(8\pi\right)^{-1}$ to be a thermodynamical variable as well. Accordingly, the first law of thermodynamics can be obtained as
\be \label{dMh}
dM_h  = \frac{{\kappa _h }}{{8\pi }}dA_h  + \omega _h dN_h  + \varpi _h dl_h + \Phi _h dQ_h  + {\cal V}_h d{\cal P}_h\,,
\ee 
where the surface gravity, NUT potential, conjugate thermodynamics volume, electrostatic potential, and angular velocity at black hole horizon are given by
\be 
\kappa _h  = \frac{{r_h  - M_h  - 2{\lambda}^2 r_h \left( {r_h  + 3l^2 } \right)}}{{\Xi _h }}\,,
\ee 
\be 
\varpi _h  = 2lr_h  - \frac{{1 + 2{\lambda}^2 \left( {r_h^2  - 3l^2 } \right)}}{{\Xi _h }}\,,
\ee 
\be 
{\cal V}_h = - \frac{{4\pi r_h \left( {r_h^4  + 6l^2 r_h^2  - 3l^4 } \right)}}{{3\Xi _h }}\,,
\ee 
\be \label{PhiH}
\Phi _h  = \frac{{Q_h r_h }}{{r_h^2  + l^2 }}\,,
\ee 
and
\be 
\omega _h  = \frac{l}{{\Xi _h }}\,,
\ee 
respectively. One can understand that eq. (\ref{dMh}) reflects the conservation of black hole energy where it now incorporates the variations in NUT parameter $l_h$ and the conserved quantity $N_h$ that associates to it.  Furthermore, using the results above, the following Bekenstein-Smarr mass formula can be verified as
\be \label{SmarrH}
M_h  = 2T_h S_h  + 2\omega _h N_h  + \varpi _h l_h + \Phi _h  Q_h  - 2{\cal V}_h {\cal P}_h\,,
\ee
where the Hawking temperature is given by $T_h = \kappa_h / 2\pi$. In the last equation, the physical parameters $M_h$, $N_h$, $l_h$, $Q_h$, and $P_h$ are the fixed conserved quantities measured at cosmological horizon as the boundary which are related to the quantities $T_h$, $S_h$, $\omega _h$, $\varpi _h$, $ \Phi _h$, and ${\cal V}_h$ evaluated at the black hole horizon.

\subsection{Cosmological horizon thermodynamics}

Now let us turn to the thermodynamical aspects associated the cosmological horizon. The prescription is similar to the previous one on the black hole horizon thermodynamics, just we need to switch the role of horizons. If in the previous subsection we considered that cosmological horizon is just a boundary, here the black hole horizon which now plays the role as a boundary. The incorporated physical parameters are
\[
M_c =- M~~,~~N_c =- Ml~~,~~Q_c=-Q~,
\]
\be 
l_c = l~~,~~\Lambda_c =- 3{\lambda}^2\,.
\ee 
From the last equation, we can have $N_c = - N_h$ as an implication that the NUT parameter does not change sign as one switch the horizons. Eventually, it has a significant impact when we try to express the relation between first laws of thermodynamics that correspond to each horizon in the spacetime. 

The area of cosmological horizon in RNTNdS spacetime (\ref{RNTNdSmetric}) is given by $A_c = 4\pi \Xi_c$ where $\Xi_c = r_c^2 +l^2$. Likewise, the condition $\Delta_Q\left(r_c\right)=0$ can give us
\be 
\Xi _c  = 2M r_c  + 2l^2  - Q_c^2  + {\lambda}^2 \left( {r_c^4  + 6l^2 r_c^2  - 3l^4 } \right)\,,
\ee 
and similar to the black hole horizon case we can have
\be 
4{M_c}^2 \Xi _c  = 4N_c^2  + \left\{ {\left( {1 - 4{\lambda}^2 l^2 } \right)\left( {\Xi _c  - 2l^2 } \right) + Q^2  - {\lambda}^2 \Xi _c^2 } \right\}^2 \,.
\ee 
An arrangement to the last equation can produce
\be \label{MassC}
{M_c}^2  = \frac{1}{{4\Xi _c }}\left\{ \left( {1 - \frac{{32\pi }}{3}{\cal P}_c l^2 } \right)\left( {\Xi _c  - 2l^2 } \right) + Q_c^2 \right.
\left.  - \frac{{8\pi }}{3}{\cal P}_c \Xi _c^2  \right\}^2  + \frac{{N_c^2 }}{{\Xi _c }}\,,
\ee
where the thermodynamical variable ${\cal P}_c=\Lambda_c \left(8\pi\right)^{-1}$ is the generalized pressure that corresponds to the cosmological horizon. 

Consequently, the mass formula (\ref{MassC}) corresponds to the first law of thermodynamics 
\be \label{dMc}
d M_c  = \frac{{\kappa _c }}{{8\pi }}dA_c  + \omega _c dN_c  + \varpi _c dl_c + \Phi _c dQ_c  + {\cal V}_c d{\cal P}_c\,,
\ee 
where the surface gravity, NUT potential, conjugate thermodynamics volume, electrostatic potential, and angular velocity at cosmological horizon are given by
\be 
\kappa _c  = \frac{{r_c  - M_c  - 2{\lambda}^2 r_c \left( {r_c  + 3l^2 } \right)}}{{\Xi _c }}\,,
\ee 
\be 
\varpi _c  = 2lr_c  - \frac{{1 + 2{\lambda}^2 \left( {r_c^2  - 3l^2 } \right)}}{{\Xi _c }}\,,
\ee 
\be 
{\cal V}_c =- \frac{{4\pi r_c \left( {r_c^4  + 6l^2 r_c^2  - 3l^4 } \right)}}{{3\Xi _c }}\,,
\ee
\be \label{PhiC}
\Phi _c  = -\frac{{Q_c r_c }}{{r_c^2  + l^2 }}\,,
\ee
and 
\be 
\omega _c  = \frac{l}{{\Xi _c }}\,,
\ee
respectively. If eq. (\ref{dMh}) can be interpreted as the law of conservation energy related to the black hole horizon, then eq. (\ref{dMc}) reflects the conservation of energy inside the cosmological horizon. Since the spacetime possesses a NUT parameter $l_c$, the total energy inside cosmological horizon gets some contributions from the changes of rotation-like and electromagnetic-like energies that come from this NUT parameter. Using the results above, the generalized Smarr mass formula
\be 
{M_c}  = 2T_c S_c  + 2\omega _c N_c  + \varpi _c l + \Phi _c Q_c  - 2{\cal V}_c {\cal P}_c
\ee 
can be justified. This result tells us the total energy inside the cosmological horizon that can be accessed by an observer outside the black hole horizon. 

As the author of \cite{Sekiwa:2006qj} has showed, the relations between physical parameters that associate to each horizon allow one to connect the corresponding first law of thermodynamics. By using the equations $dM_h = - dM_c$, $dN_h = - dN_c$, $dl_h = dl_c$, $dQ_h = -dQ_c$, and $d{\cal P}_h = -d{\cal P}_c$, adding eq. (\ref{dMh}) to eq. (\ref{dMc}) can give us
\be \label{ConsvTotal}
\left( {{\cal V}_{h}  - {\cal V}_{c} } \right)d\left( { {\cal P}_{c} } \right) + \left( {\Phi _{h}  - \Phi _{c} } \right)d\left( { Q_{c} } \right) + \left( {\omega _{h}  - \omega _{c} } \right)d\left( { N_{c} } \right) - \left( {\varpi _{h}  + \varpi _{c} } \right)d\left( {  l_{c} } \right) - T_h dS_h  = T_c dS_c\,.
\ee 
Basically, eq. (\ref{ConsvTotal}) is similar to the result presented in ref. \cite{Sekiwa:2006qj} for the black hole horizon and cosmological horizon entropy relation in the Kerr-Newman-de Sitter spacetime
\be \label{gen.thermo.relation.ori}
\left( {{\cal V}_c  - {\cal V}_h } \right)d\left( { - {\cal P}_c } \right) + \left( {\Phi _c  - \Phi _h } \right)d\left( { - Q_c } \right) + \left( {\Omega _c  - \Omega _h } \right)d\left( { - J_c } \right)
- T_h dS_h  = T_c dS_c \,,
\ee 
except the term that contains the sum of NUT potentials $\varpi _c$ and $\varpi _h$. Note that the conserved angular momentum $J_h = - J_c$ in eq. (\ref{gen.thermo.relation.ori}) corresponds to the rotational energies in the rotating spacetime. On the other hand, the quantity $N_h = - N_c$ in eq. (\ref{ConsvTotal}) exists due to the presence of NUT parameter in the spacetime which can be interpreted due to the existence of Misner string attached to the south and north poles. In some other works, the authors even show how to associate this Misner string some entropy which then can be incorporated in the corresponding first law of thermodynamics \cite{BallonBordo:2019vrn}.

To extract the physical meaning of eq. (\ref{ConsvTotal}), let us compare it to the similar equation as given in (\ref{gen.thermo.relation.ori}). It is easy to observe that the first three terms in the l.h.s. of both equations are analogous. The first term reflects the increase of vacuum energy inside the cosmological horizon, the second one describes the extracted electromagnetic energy\footnote{Note that the RNTNdS spacetime possesses the electric charge $Q$ as well.}, and the third term represents the extracted rotational or rotational-like energy. However, the fourth term of l.h.s in eq. (\ref{ConsvTotal}) has no analogous counterpart in the first law of Kerr-Newman-de Sitter black hole termodynamics (\ref{gen.thermo.relation.ori}). We interpret this fourth term as total electromagnetic-like energy inside the cosmological horizon that comes from the presence of Misner string which extends in the visible region $r_h < r < r_c$, i.e. from the black hole surface to the cosmological horizon. Finally the fifth term in l.h.s. of eq (\ref{ConsvTotal}), and also the fourth term in l.h.s. of eq. (\ref{gen.thermo.relation.ori}), is related to the Hawking radiation of the black hole. From eq. (\ref{ConsvTotal}) we can learn that the energy decrease in the visible region and the growing of black hole mass would lead to the decreasing of cosmological horizon entropy. 

Before coming to the conclusion, let us add some remarks here on the obtained results in this section. Indeed, the Bekenstein-Smarr mass formula (\ref{SmarrH}) can also apply for the horizon of \RN-Taub-NUT-anti de Sitter (RNTNAdS) black hole as discussed in \cite{Wu:2019pzr}. It can be performed by replacing $\lambda^2 \to - \lambda^2$ in the corresponding equations. Particularly, the generalized pressure ${\cal P}_h$ changes sign in equation (\ref{dMh}) which reflects the different way for the total mass to change with respect to the variation of generalized pressure ${\cal P}_h$. Surely the RNTNAdS spacetime does not solve the equations of motion in the DGP brane presented in section \ref{sec.2.effectiveeqtn3brane}, as it corresponds to the Einstein equations with a negative cosmological constant. Moreover, cosmological horizon does not exist in the anti-de Sitter spacetime. Therefore, the results obtained above for cosmological horizon thermodynamics and the related relation in eq. (\ref{ConsvTotal}) that connects two entropies do not apply for the RNTNAdS spacetime.

\section{Conclusion}\label{sec.conclusion}

In this paper, we have showed that the RNTNdS spacetime of Einstein-Maxwell theory solves exactly the non-flat case four dimensional equations of motion in DGP braneworld scenario, provided a particular form of the five dimensional Weyl tensor projection is satisfied. This approach in solving equations of motion in a braneworld theory has been applied to several cases, including a recent one in RS-II brane \cite{Siahaan:2020bga}. However, unlike the spacetime solution in RS-II braneworld that can include the so called tidal charge, the Hamiltonian constraint conditions in DGP theory cannot allow such charge to exist. 

The thermodynamics of RNTNdS spacetime on the DGP brane was discussed in section \ref{section.5.thermo}. The positive cosmological constant is understood to be related with the cross-over scale in the theory. By adopting the approach by Sekiwa \cite{Sekiwa:2006qj} in discussing the black hole thermodynamics in de Sitter spacetime, and the proposal by Wu et al. \cite{Wu:2019pzr} in defining a new conserved quantity that associates to NUT parameter, we managed to write down the generalized Smarr formula and the first law of thermodynamics for each horizon in the RNTNdS spacetime. Interestingly, the first law of thermodynamics for each horizon can be combined to give an equation which describes the change of cosmological horizon entropy with respect to the growth of black hole mass and the energies in the visible region. This result is novel as there is no previous works have been addressed to study the thermodynamics of a Taub-NUT de Sitter spacetime in this fashion. 

However, we have not considered the Gibbs energy that associates to the spacetime in this paper. As it has been shown in ref. \cite{Kubiznak:2015bya}, the Gibbs energy can be an effective tool to study the thermodynamics of a de Sitter spacetime. Moreover, we can also add the inner horizon entropy that can contribute to the total first law of thermodynamics in the system. Incorporating rotation can also be an interesting addition to the solution finding as worked out in section \ref{section.4.charged} or to the thermodynamics investigation as presented in section \ref{section.5.thermo}. Moreover, insight from the AdS/CFT correspondence has shown that considering the cosmological constant as one of the thermodynamic variables is not the correct way to obtain the laws of spacetime thermodynamic \cite{Papadimitriou:2005ii}. To do so, one should proceed by using the Noether-Wald method \cite{Wald:1993nt} which relies on the asymptotic charges. These problems can be included in our future work.

\section*{Acknowledgement}

This work is supported by Lembaga Penelitian dan Pengabdian kepada Masyarakat Universitas Katolik Parahyangan under contract no. III/LPPM/2022-02/79-P. I thank the reviewers for his/her comments on the manuscript.


\begin{thebibliography}{99}
	
	\bibitem{Nojiri:2003rz}
	S.~Nojiri and S.~D.~Odintsov,
	Phys. Lett. B \textbf{576} (2003), 5-11
	
	\bibitem{Hou:2021okc}
	Y.~Hou, M.~Guo and B.~Chen,
	Phys. Rev. D \textbf{104} (2021) no.2, 024001
	
	\bibitem{Eiroa:2017uuq}
	E.~F.~Eiroa and C.~M.~Sendra,
	Eur. Phys. J. C \textbf{78} (2018) no.2, 91
	
	\bibitem{Dey:2020lhq}
	R.~Dey, S.~Chakraborty and N.~Afshordi,
	Phys. Rev. D \textbf{101} (2020) no.10, 104014
	
	\bibitem{Toshmatov:2016bsb}
	B.~Toshmatov, Z.~Stuchl\'\i{}k, J.~Schee and B.~Ahmedov,
	Phys. Rev. D \textbf{93} (2016) no.12, 124017
	
	\bibitem{hor}
	P. Horava and E. Witten, 
	Nucl. Phys. B {\bf460} (1996) 506
	
	\bibitem{ark}
	N. Arkani-Hamed, S. Dimopoulos and G. Dvali, 
	Phys. Lett. B {\bf429} (1998) 263
	
	\bibitem{ran}
	L. Randall and R. Sundrum, 
	Phys. Rev. Lett. {\bf83} (1999) 4690
	
	\bibitem{dgp}
	G. Dvali, G. Gabadadze and M. Porrati, 
	Phys. Lett. B {\bf485} (2000) 208
	
	\bibitem{Gholami:2021olb}
	F.~Gholami, F.~Darabi and A.~H.~Badali,
	Indian J. Phys. \textbf{96} (2022) no.3, 963-969
	
	
	\bibitem{Iqbal:2019ooy}
	A.~Iqbal and A.~Jawad,
	Phys. Dark Univ. \textbf{26} (2019), 100349
	
	
	\bibitem{Chetry:2019fhz}
	B.~Chetry, J.~Dutta, U.~Debnath and W.~Khyllep,
	Int. J. Geom. Meth. Mod. Phys. \textbf{16} (2019) no.11, 1950173
	
	\bibitem{Warkentin:2019caf}
	M.~Warkentin,
	JHEP \textbf{03} (2020), 015
	
	\bibitem{Biswas:2018mxe}
	M.~Biswas, S.~Ghosh and U.~Debnath,
	Int. J. Geom. Meth. Mod. Phys. \textbf{16} (2019) no.11, 1950178
	
	\bibitem{Sbisa:2018ydq}
	F.~Sbis\`a,
	Universe \textbf{4} (2018) no.12, 136
	
	\bibitem{Jawad:2018zeo}
	A.~Jawad and A.~Iqbal,
	Eur. Phys. J. Plus \textbf{133} (2018) no.11, 470
	
	\bibitem{Chang-Young:2007iiv}
	E.~Chang-Young and D.~Lee,
	Phys. Lett. B \textbf{659} (2008), 58-64
	
	\bibitem{Lee:2007nk}
	D.~Lee, E.~Chang-Young and M.~Yoon,
	Int. J. Mod. Phys. A \textbf{24} (2009), 4389-4401
	
	\bibitem{Lee:2008av}
	D.~Lee, E.~Chang-Young and M.~Yoon,
	Phys. Lett. B \textbf{663} (2008), 11-16
	
	\bibitem{Griffiths:2009dfa}
	J.~B.~Griffiths and J.~Podolsky,
	\textit{Exact Space-Times in Einstein's General Relativity}, (Cambridge University Press, Cambridge, 2009)
	
	\bibitem{Zhou:2022eiv}
	C.~Zhou,
	Eur. Phys. J. C \textbf{82} (2022) no.10, 886
	
	\bibitem{Liu:2022wku}
	H.~S.~Liu, H.~Lu and L.~Ma,
	JHEP \textbf{10} (2022), 174
	
	\bibitem{Awad:2022jgn}
	A.~Awad and S.~Eissa,
	Phys. Rev. D \textbf{105} (2022) no.12, 124034
	
	\bibitem{Gustavsson:2022jpo}
	A.~Gustavsson,
	JHEP \textbf{09} (2022), 153
	
	\bibitem{Siahaan:2020bga}
	H.~M.~Siahaan,
	Phys. Rev. D \textbf{102} (2020) no.6, 064022
	
	\bibitem{Dadhich} N. Dadhich, R. Maartens, P. Papadopoulos and V. Rezania,
	Phys. Lett. B {\bf 487}, 1 (2000) 
	
	\bibitem{Aliev:2007fy}
	A.~N.~Aliev,
	Phys. Rev. D \textbf{77}, 044038 (2008)
	
	\bibitem{Teitelboim:2002cv}
	C.~Teitelboim,
	arXiv:hep-th/0203258.
	
	\bibitem{Sekiwa:2006qj}
	Y.~Sekiwa,
	Phys. Rev. D \textbf{73} (2006), 084009
	
	\bibitem{BallonBordo:2019vrn}
	A.~Ballon Bordo, F.~Gray, R.~A.~Hennigar and D.~Kubiz\v{n}\'ak,
	Phys. Lett. B \textbf{798} (2019), 134972
	
	
	\bibitem{Hennigar:2019ive}
	R.~A.~Hennigar, D.~Kubiz\v{n}\'ak and R.~B.~Mann,
	Phys. Rev. D \textbf{100}, no.6, 064055 (2019)
	
	
	\bibitem{Wu:2019pzr}
	S.~Q.~Wu and D.~Wu,
	Phys. Rev. D \textbf{100} (2019) no.10, 101501
	
	\bibitem{Aliev:2004ds}
	A.~Aliev and A.~Gumrukcuoglu,
	Class. Quant. Grav. \textbf{21}, 5081-5096 (2004)
	
	\bibitem{Chamblin}
	A. Chamblin, S. W. Hawking and H. S. Reall, 
	Phys. Rev. D {\bf61} (2000) 065007
	
	\bibitem{Shiromizu}
	T.~Shiromizu, K.~i.~Maeda and M.~Sasaki,
	Phys. Rev. D \textbf{62} (2000), 024012
	
	\bibitem{Gomberoff:2003ea}
	A.~Gomberoff and C.~Teitelboim,
	Phys. Rev. D \textbf{67} (2003), 104024
	
	\bibitem{Caldarelli:1999xj}
	M.~M.~Caldarelli, G.~Cognola and D.~Klemm,
	Class. Quant. Grav. \textbf{17} (2000), 399-420
	
	\bibitem{Kubiznak:2015bya}
	D.~Kubiznak and F.~Simovic,
	Class. Quant. Grav. \textbf{33} (2016) no.24, 245001
	
	\bibitem{Lee:1990nz}
	J.~Lee and R.~M.~Wald,
	J. Math. Phys. \textbf{31} (1990), 725-743
	
	\bibitem{Wald:1993nt}
	R.~M.~Wald,
	Phys. Rev. D \textbf{48} (1993) no.8, R3427-R3431
	
	\bibitem{Anninos:2010zf}
	D.~Anninos, G.~S.~Ng and A.~Strominger,
	Class. Quant. Grav. \textbf{28} (2011), 175019
	
	\bibitem{Compere:2020lrt}
	G.~Comp\`ere, A.~Fiorucci and R.~Ruzziconi,
	JHEP \textbf{10} (2020), 205
	
	\bibitem{Kolanowski:2020wfg}
	M.~Kolanowski and J.~Lewandowski,
	Phys. Rev. D \textbf{102} (2020) no.12, 124052
	
	\bibitem{Poole:2021avh}
	A.~Poole, K.~Skenderis and M.~Taylor,
	[arXiv:2112.14210 [hep-th]]
	
	
	\bibitem{Mann:2005cx}
	R.~B.~Mann and C.~Stelea,
	Phys. Lett. B \textbf{634} (2006), 531-535
	
	\bibitem{Papadimitriou:2005ii}
	I.~Papadimitriou and K.~Skenderis,
	JHEP \textbf{08} (2005), 004
	
\end{thebibliography}
\end{document}